# Taiji Patterns Swirled Out of Magnetic Liquid Metal Fluids under Rotating Magnets

Wentao Xiang[a,b], and Jing Liu[a,b*]

[a] *State Key Laboratory of Cryogenic Science and Technology, Technical Institute of Physics and Chemistry, Chinese Academy of Sciences, Beijing 100190, China*
[b] *School of Future Technology, University of Chinese Academy of Sciences, Beijing 100049, China*
[*]*Corresponding author. E-mail: jliu@mail.ipc.ac.cn*

**Abstract**

While magnetic fluids are well known for their rich ferrohydrodynamic behaviors, prior researches on dynamic droplet morphologies have largely been confined to the nonconducting matters. From an alternative, the electrically conductive magnetic liquid metals offer ever larger space for explorations meanwhile also incubate intriguing mysteries that had not been understood before. Here, we disclosed a group of rather profound fluidic phenomena happening on the magnetic liquid metals with sizes spanning from millimeter to centimeter scales when subject to the impact of a rotating magnet pair. We conceived that the identified flow patterns highly resemble that of the Chinese Taiji diagrams widely known as a classical theory to describe all things genesis and transformation, indicating the kaleidoscopic variations and intrinsic clues between modern magnetic liquid metal physics and the nature rules lying behind the ancient Eastern philosophy. Through systematically tuning the droplet volume, magnetic field strength, and magnet rotational speed, we achieved precise control over the liquid metal fluidic morphologies such as ellipses, dumbbells, toroidal rings, and Yin-Yang symbols, and quantified their dependence on the prescribed experimental conditions. Following the routes to construct the Taiji eight diagrams and their derivatives, we classified the disclosed flow patterns into eight representative schemes and three phase diagrams were plotted to characterize the fluidic patterns of the magnetic droplets correspondingly. The underlying mechanisms of magnetic liquid metal flow separation and coalescence were further interpreted from an energy perspective. These findings suggest enormous experimental insights and a theoretical framework for advancing fundamental magnetohydrodynamics and related engineering practices.

**Keywords**: magnetic fluids, liquid metals, rotating magnetic field, magnetic droplet, flow pattern

## 1. Introduction

Magnetic fluids are suspensions consisting of nanosized or micronsized magnetic particles homogeneously dispersed within a carrier liquid. Conventionally, these fluids employ water, oils, or organic solvents as the carrier liquid. They exhibit a unique combination of inherent magnetic responsiveness and hydrodynamic behavior, leading to a variety of complex field-induced phenomena. Notably, under the influence of rotating magnetic fields, nonconducting magnetic fluids would rotate with the magnetic field [1,2]. This rotational effect may be caused by vortex diffusion induced by the rotation of magnetic particles within the magnetic fluids [3], or through magnetic field gradient distribution resulting from heat generation due to particle rotation [4]. Additionally, magnetic fluid microdroplets (with a typical radius of 10 μm) are found to undergo distinctive morphological transformations under a rotating magnetic field, giving rise to structures such as wormlike configurations and multi-spiked, starfish-like formations [5–7].

Replacing non-conductive carrier liquids with room-temperature liquid metals (LMs) imparts high electrical conductivity to LM-based magnetic fluids (LMMFs), significantly broadening their functional scopes compared to conventional magnetic fluids. The commonly used room-temperature LMs generally include mercury, sodium–potassium alloys, and gallium-based alloys, among others. Earlier studies have successfully synthesized ferrofluids using mercury and sodium–potassium alloys [8–10]. However, serious concerns over their toxicity or high chemical reactivity have limited their extensive research and practical applications. In recent years, with the expanding applications of gallium-based LMs and advances in preparation techniques for gallium-based LMMFs, metallic magnetic fluids have regained considerable research interest. Owing to their high electrical conductivity, distinctive interfacial behavior, and tunable chemical properties, the gallium-based LMs flow can be precisely controlled through various external stimuli—such as electric fields, electromagnetic actuation, and electrochemical methods—leading to diverse flow phenomena and pattern formations [11]. Regarding the magnetic induction issue, the behavior of pure LMs under rotating magnetic fields had been studied before. According to Faraday's law, time-varying magnetic fields would induce eddy currents within conductive LMs, generating electromagnetic body forces that drive a range of fluid dynamic responses. These include bulk annular flow [12–14], orbital motion of LM droplets [15,16], and electromagnetic levitation of LM droplets [17]. Such effects have been even harnessed in some cutting edge applications such as pumps and reconfigurable antennas [18].

When integrated with the magnetism of suspended magnetic particles, LMMFs will exhibit more intricate physical behaviors [19]. Studies have confirmed that these fluids display characteristic magnetic responses and electromagnetic induction phenomena. For instance, under static magnetic fields, LMMF droplets avoid the formation of annular waves or curled edges upon impact [20] and demonstrate notable magnetorheological effects [21]. Under alternating magnetic fields, the magnetothermal effect [22] and magnetic susceptibility variations [23] in LMMFs have also been investigated and utilized. Nevertheless, research on the behavior of LMMFs under rotating magnetic fields remains limited. Previous studies have focused on the deformation characteristics of micro-sized LMMF droplets (average size from ~200 nm to ~2 μm) under magnetic fields generated by rotating individual bar magnets [24,25]. These studies revealed that the droplets develop spikes, which can be utilized for medical sterilization.

In this work, we are dedicated to disclose the fluidic and transformation phenomena of LMMF droplets under a rotating magnetic field. Using a magnetic stirrer, we observed a very wide range of flowing patterns of the LMMF droplets, such as spin, torus flow, Taiji-like flow, and more, to name only a few, while their positions remain fixed by the magnetic confinement. It occurs to us that the observed fluidic phenomena resemble various complex patterns predicted by the eight diagrams and their derivatives in ancient Chinese Yin-Yang philosophy. During the spin process, the LMMF droplet morphologies may transform from ellipsoidal to dumbbell-shaped as the magnetic field rotation speed increases, and subsequently undergo repeated splitting and coalescence. Under specific conditions, the experiment also allows for the observation of multiple droplets moving along separate trajectories, similar to the motion of galaxies or satellites. Some of these phenomena could not be observed in conventional magnetic fluids otherwise. We attribute them to the coupled effects of magnetohydrodynamics of LMs, high surface tension of LMs, and the magnetic properties of the magnetic particles. And the intrinsic clues of such unusual fluidic activities to the Classical

Taiji diagram were preliminarily disclosed.

## 2. Principles and Methods
### 2.1. Experimental Methods

As shown in Figure 1(a), the present LMMF was prepared by thoroughly mixing LM and iron powder in a hydrochloric acid environment (~1 mol L$^{-1}$) with a mass ratio of iron powder to LM of 1:9. The hydrochloric acid can remove surface oxides from both the LM and iron powder, enhancing their mutual wettability [21,26]. The LM selected is EGaIn (eutectic gallium indium, 75.5 wt.% Ga and 24.5 wt.% In), characterized by its melting point of 15.5 °C, density of 6.28 g cm$^{-3}$, surface tension of 0.624 N m$^{-1}$, and electrical conductivity of 3.45×10$^6$ S m$^{-1}$ [19]. The iron powder has an average particle size of 10 μm. A rotating magnetic field was generated by a magnetic stirrer. This stirrer contained a pair of permanent magnets (magnet diameter 25 mm, spaced 40 mm apart), driven by an electric motor to rotate clockwise and produce the rotating magnetic field. Figure 1(b) and (c) show the schematic and photograph of the experimental setup, respectively. The LMMF flow effect exhibits a strong correlation with droplet volume and magnetic field strength. In the experiment, the volume of the magnetic droplet was controlled using a pipette gun. Magnetic field strength was regulated by adjusting the distance, z, between the container and the substrate. The z-direction distance was increased by progressively adding more 3 mm thick acrylic plates (non-metallic, non-magnetic material) at the bottom. The camera used for photographing and video recording in the experiment is a digital single-lens reflex camera (EOS 90D, Canon, Japan), with a frame rate of 50 fps.

We confirmed the magnetic field distribution above the magnet through a combination of simulation and actual measurement. By testing the magnetic induction intensity at the center and above the magnet at different distances from the magnet pair using a digital Tesla meter (KT-102, KeOuTe, China), we determined the parameters for the simulated magnet pair. The COMSOL simulation results of the magnetic field distribution are shown in Figure 1(d). Specifically, at different distances above the magnet, the magnetic flux density in the z-direction ($B_z$) and the magnetic flux density in the x-direction ($B_x$) exhibit markedly different distribution functions. $B_z$ exhibits an axisymmetric distribution along the axis, with equal magnetic induction magnitudes on both sides of the axis but in opposite directions (Figure 1(e)). $B_x$ exhibits a symmetrical distribution along the axis, with magnetic induction of equal magnitude and identical direction on both sides (Figure 1(f)). Simultaneously, a trough in magnetic field strength appears near the magnet pair, gradually diminishing with increasing distance. These magnetic field characteristics exert a critical influence on the fluid behavior of the LMMF droplet. A schematic of the spin of a single LMMF droplet and the external field in the experiment is shown in Figure 1(g).

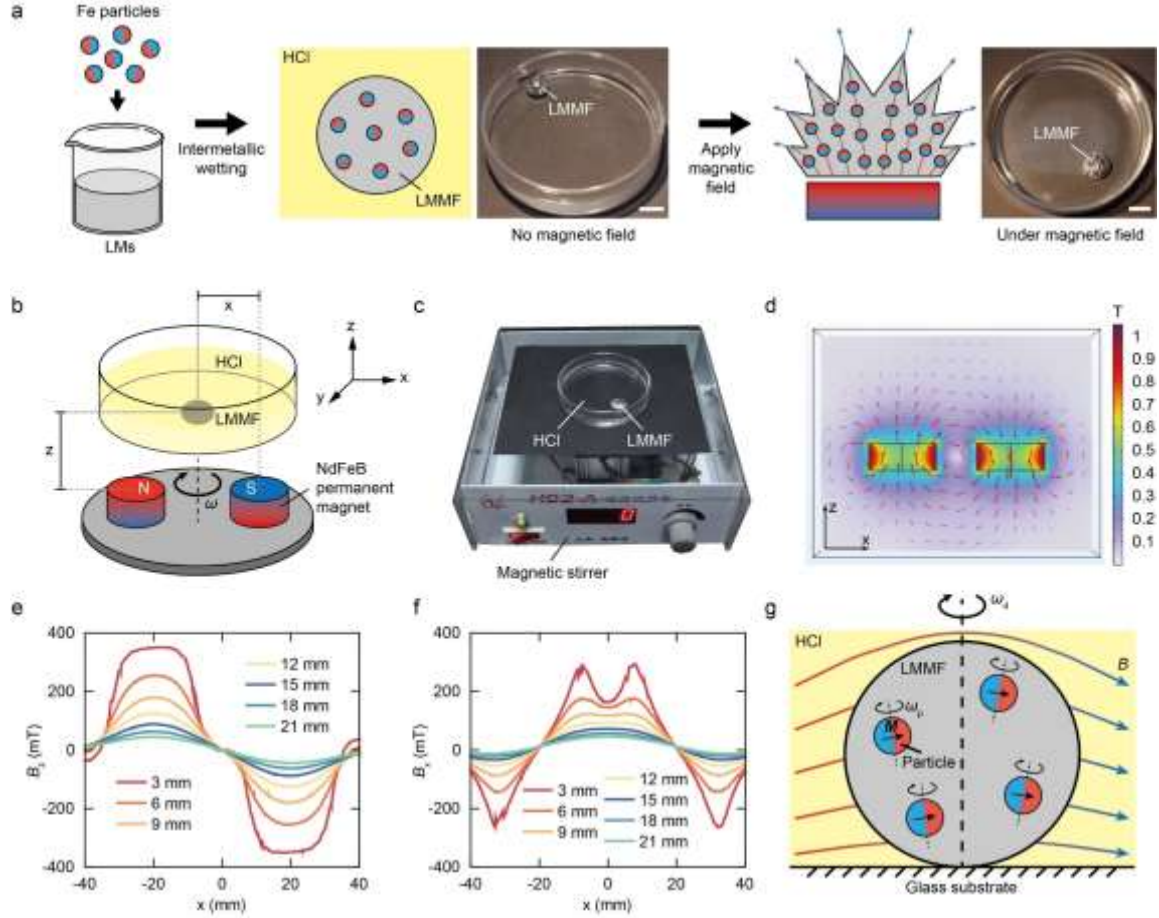

Figure 1. Experimental setup and conditions. (a) Preparation and photographs of LMMF. Gallium-based LM mixed with Fe particles forms LMMF in a hydrochloric acid environment due to intermetallic wetting. Without a magnetic field, the droplet shape resembles that of a pure LM droplet. Upon application of a static magnetic field, small spikes emerge on the droplet surface. Scale bar：1 cm. (b) Schematic of the experimental setup. The rotating magnetic field is generated by a motor driving a pair of cylindrical NdFeB magnets to rotate. (c) Photograph of the experimental setup. (d) Side view of the simulated magnetic field distribution generated by the magnet pairs in the experiment. (e) Simulation results of the distribution function of the z-axis magnetic field $B_z$ strength along the x-axis at different distances (z = 3-21 mm) directly above the center of the magnet. (f) Simulation results of the distribution function of the x-axis magnetic field strength along the x-axis ($B_x$) at different distances (z = 3-21 mm) directly above the center of the magnet. (g) Flow analysis of LMMF droplet. LMMF droplets spin due to the rotation of magnetic particles and LM.

## 2.2. Kaleidoscopic LMMF Pattern and Its Taiji Theory Implication

Aiming to systematically tackle the main target issues, we conducted experiments on LMMF droplets with three different volumes (500 μL, 300μL, and 100 μL). The fluidic behavior of each droplet was recorded and tested under varying magnetic field strengths (i.e., varying distance from magnet pair, z = 3-21 mm) and field rotational speeds ($\omega$ = 200-1930 rpm). A series of unusual fluidic phenomena were thus disclosed.

Initial observations focus on the fluid behavior of larger LMMF droplets (500 μL) under weak magnetic fields. Upon energizing the magnetic stirrer, the LMMF droplet becomes fixed at the center of the magnet pair and executes a steady spin, during which it deforms into an ellipsoid (Figure 2(a-i)). The spin angular velocity of the droplet and its aspect ratio ($a/c$) are found to increase with field rotational speed. While the droplet spin remains synchronous with the magnetic field at low speeds, the onset of asynchronous motion is detected beyond a critical velocity, concomitant with a shape transition to a dumbbell-like structure comprising two ellipsoidal termini linked by a fine filament (Figure 2(a-ii)). Subsequent acceleration induces substantial centrifugal forces, leading to either the ejection of poorly-wetting LM droplets or a cyclic separation-coalescence dynamic (Figure 2(a-iii)). This repetitive cycle involves the centrifugal separation of the droplet into two subunits, their re-coalescence via collision and centripetal action, and subsequent re-separation. These morphological evolutions are governed by the energy balance between surface tension and magnetic field.

As the magnetic field intensifies, the correlation between droplet dynamics and magnetic field rotational speed undergoes a marked transformation. At low rotational speeds, the LMMF droplet exhibits a growing tendency to synchronize with the bottom rotating magnet (Figure 2(b-i)), rather than spinning at the center of the magnet pair. There exists a threshold velocity at which droplets transition from synchronous magnetic motion to spin at the center. Beyond a certain threshold velocity, the droplet undergoes a change in shape at the magnet pair center. At larger field rotational speeds, droplet separation becomes highly irregular, qualitatively characterized by an increased number of satellite droplets (Figure 2(b-ii) and 2(b-iii)). As the magnetic field speed continues to increase, the satellite droplets diminish, and the droplets tend to undergo complex, twisted deformations around the center of the magnet pair (Figure 2(b-iv) and 2(b-v)). Under the strongest magnetic field condition ($z = 3$ mm), the droplet dynamics display a more complex behavioral regime. With increasing field rotational speed, the LMMF droplet splits, and the resulting daughter droplets follow divergent trajectories, inducing vigorous agitation in the surrounding hydrochloric acid medium (Figure 2(c-ii)). Beyond a threshold velocity, the droplet no longer vigorously stirs the hydrochloric acid solution and exhibits a quasi-ring motion pattern (Figure 2(c-iii) and 2(c-iv)). At $\omega = 1400$ rpm, LMMF torus forms—a configuration that resembles the planetary ring dynamics theorized by Poincaré (Figure 2(c-v)). A coupling emerges between the intrinsic tendency for droplet separation and the magnetic centripetal effect, leading the droplet to develop this toroidal flow tendency. To our surprise, with further increase in field rotational speed, the coherent ring structure eventually breaks apart, giving rise to an outstanding flow pattern completely resembling the classical Taiji pattern, named for the distinctive Yin–Yang morphology formed by its bifurcated droplets, evocative of the traditional Chinese Taiji symbol (Figure 2(c-viii)).

Further test is conducted on the behavior of 300 μL droplets. Under a weak magnetic field, it is observed that the droplet rotates synchronously while elongating with increasing field rotational speed (Figure 2(d)). A critical threshold marks the onset of a cyclic separation-coalescence process. This separation process is suppressed at higher speeds, where the droplet instead maintains a stable, rotating dumbbell configuration. Enhancement of the magnetic field is found to increase both the droplet angular velocity and its propensity for separation, resulting in a larger number of polydisperse daughter droplets (Figure 2(e)). Notably, under moderate magnetic field strength and low magnetic field rotational speeds, a 300 μL LMMF droplet is observed to split into two distinct droplets exhibiting different motions—one performing spin motion and the other orbital motion (Figure 2(e-ii)). Under the strongest magnetic field ($z = 3$ mm), the LMMF droplet separates as the rotational speed increases, but they also tend towards quasi-ring motion (Figure 2(f)). However, due to the smaller droplet volume, this shape is not as pronounced as in 500μL droplets, and distinct separation becomes more irregular.

For the smallest droplet size (100 μL), under a weak magnetic field (z = 21 mm), synchronous spin of the LMMF droplet is observed at low speeds, followed by elongation into a dumbbell shape with the field rotational speed (Figure 2(g)). Separation is unlikely to occur under weak field conditions. With intensified magnetic fields and increased droplet velocity, separation commences, yielding a greater number of satellite droplets (Figure 2(h) and 2(i)). These satellite droplets also undergo a separation-coalescence process, with each droplet simultaneously spinning during the splitting phase. All these transition behaviors highly resemble the pictures revealed by the classical Taiji diagram when describing things genesis and conversions which remind us to interpret the findings via such a concise way as follows.

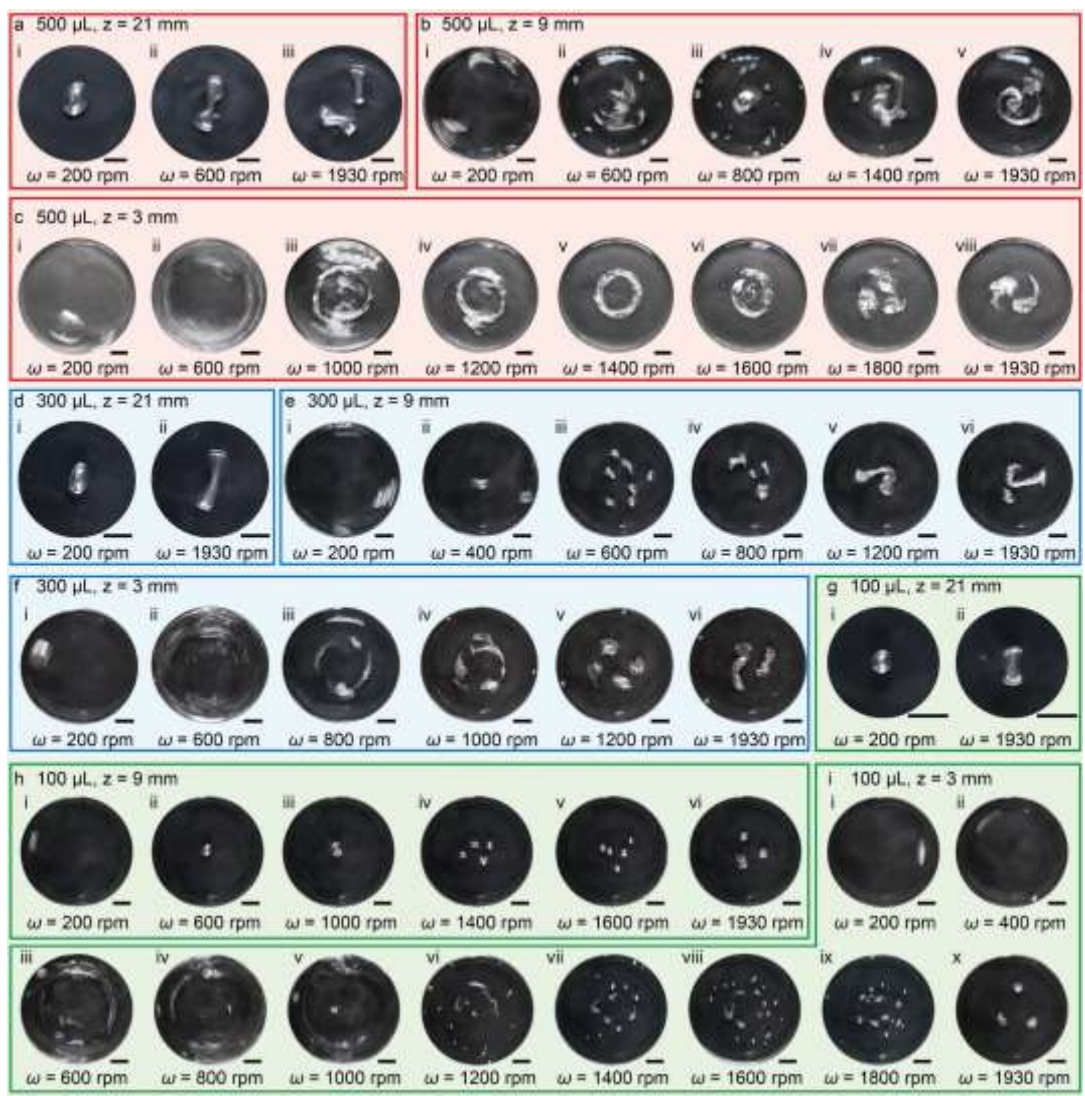

Figure 2. Typical LMMF patterns observed in the experiments. The photographed plane is the *xy* plane, with the magnetic field rotating clockwise in all cases. The red, blue, and green regions in the figure represent droplet volumes of 500 μL, 300 μL, and 100 μL, respectively. The specific conditions are as follows: (a) 500 μL LMMF with z = 21 mm. (b) 500 μL LMMF with z = 9 mm. (c) 500 μL LMMF with z = 3 mm. (d) 300 μL LMMF with z = 21 mm. (e) 300 μL LMMF with z = 9 mm. (f) 300 μL LMMF with z = 3 mm. (g) 100 μL LMMF with z = 21 mm. (h) 100 μL LMMF

with z = 9 mm. (i) 100 μL LMMF with z = 3 mm. The scale bars are all 1 cm.

**2.3. Classification of Typical LMMF Flow Patterns**

Remarkably, we discover that the fluid patterns emerging in the experiment can be analogized to the Taiji pattern. The Taiji pattern is a central symbol in ancient Chinese philosophy, originating from the *Book of Changes* (*I Ching*). Its fundamental composition comprises the central Yin-Yang symbol, the surrounding eight trigrams, and the sixty-four hexagrams derived from the outermost layer. The Yin-Yang symbol consists of interlocking black and white shapes, representing the interdependence and transformation of the dual forces of Yin and Yang. These forces cycle through growth and decline in a dynamic equilibrium, much like the separation and coalescence of LMMF droplets. Surrounding the Yin-Yang symbol are the eight trigrams, formed by the arrangement and combination of Yin and Yang lines (Yang lines refer to solid lines, while Yin lines are dashed lines). In classical Chinese philosophy, each trigram symbolizes a natural phenomenon, while the eight trigrams encapsulate the fundamental natural phenomena of the universe. In analogy and based on this, we also classify the fluid patterns emerging from LMMF under a rotating magnetic field into the eight trigrams, representing basic components of LMMF patterns, as illustrated in Figure 3(a). Corresponding actual photographs of the fluid patterns are shown in Figure 3(b). As for the outermost sixty-four hexagrams, formed by pairing and combining the eight trigrams, they represent the unfolding of phenomena over time in Chinese culture, corresponding to the temporal evolution of things. We ascribe the sixty-four hexagrams to the increasingly complex and diverse fluid patterns derived from adjusting experimental conditions.

We categorize typical fluid patterns into three phase diagrams, each corresponding to droplets of different volumes (500 μL, 300 μL, and 100 μL), as shown in Figure 4a,b,c, respectively. It can be divided into four regions based on the LMMF droplet shape. The first region is the single-droplet motion zone, typically concentrated in areas of low magnetic field strength and low field rotational speed. This includes single-droplet orbital motion, elliptical droplets, and dumbbell-shaped droplet spins. The second region is the multiple droplet zone, usually found in areas of low magnetic field strength and high field rotational speed. LMMFs split into multiple spinning droplets, undergoing periodic separation and coalescence. The third region is the unstable shape zone, typically occurring in medium magnetic fields and high rotational speeds. Drops frequently exhibit irregular deformations, displaying various distortions, splits, or serpentine motions. The fourth region is the ring or quasi-ring flow zone, typically occurring in high magnetic fields. Drops exhibit annular flow or near-annular circular motion with highly pronounced periodicity. In subsequent sections, we will conduct a detailed analysis of these phenomena.

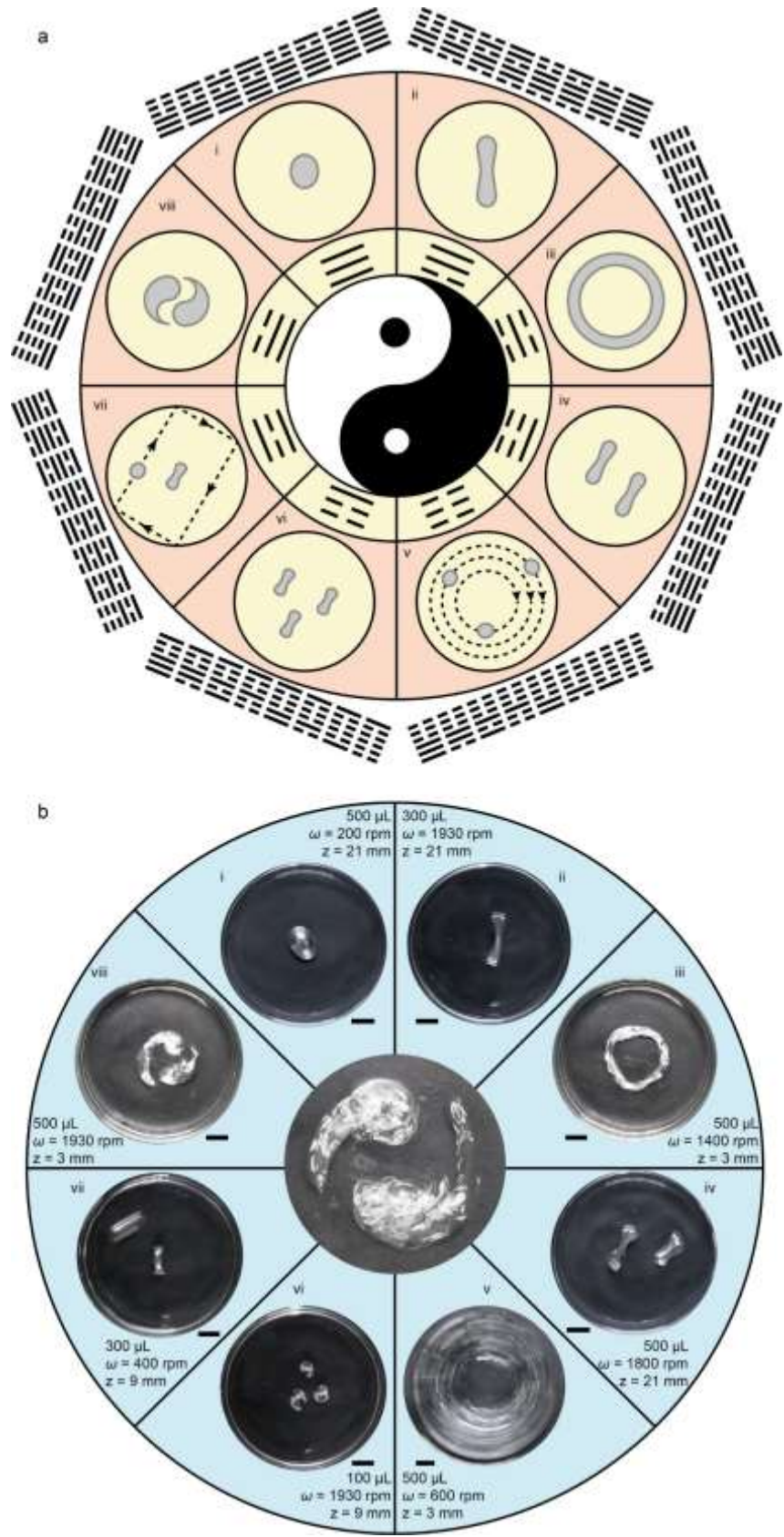

Figure 3. Analogy between typical LMMF patterns and Taiji diagrams. (a) Schematics of eight typical patterns and their corresponding experimental outputs. The most interior depicts the well-known Yin-Yang symbol, and its outside represents the eight diagrams. Further outside, eight fluidic

patterns from i to viii are classified. The patterns are: (i) Elliptical, (ii) Dumbbell-shaped, (iii) Torus-shaped, (iv) Split into two spins, (v) Split into multiple droplet moving in orbital paths, (vi) Split into three droplets performing spin motion, (vii) Split into two droplets, one spinning centrally while the other moves in a non-circular orbit, (viii) Yin-Yang pattern. These eight patterns correspond to the eight trigrams in the Chinese Taiji diagram. At the utmost, sixty-four symbols are depicted to indicate the derivatives of the eight diagrams. Through adjusting experimental conditions, more complex patterns can still be achieved, corresponding to additional trigrams (all 64 trigrams). (b) (i) to (viii) correspond to the photographs of diagrams (i) to (viii) in Figure 3(a), respectively. The scale bars in (b) are all 1 cm.

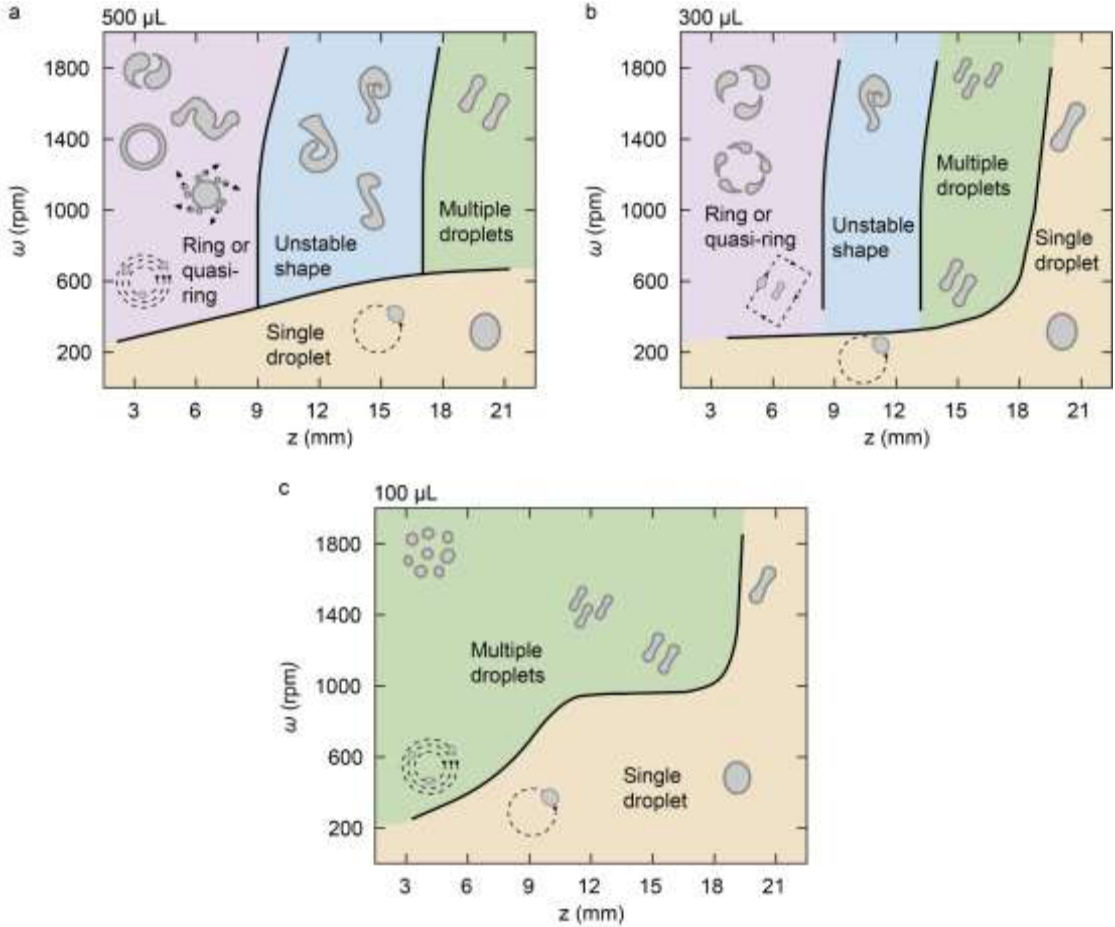

Figure 4. Phase diagram of LMMF droplet shapes. (a) Phase diagram of 500 µL LMMF droplet. (b) Phase diagram of 300 µL LMMF droplet. (c) Phase diagram of 100 µL LMMF droplet.

## 3. Theoretical Interpretation

In this section, we briefly introduce the fundamental fluid equations and droplet separation theory related to LMMF, along with a simple analysis. This would provide insights for subsequent specific analysis of fluid patterns.

### 3.1. Basic Equations for LMMFs

Previous studies have derived the fundamental fluid equations for conducting magnetic fluids based on thermodynamics [27]. Neglecting relativistic effects, the mass conservation equation for LMMFs is:

$$\frac{\partial \rho}{\partial t}+\nabla\cdot(\rho \boldsymbol{u})=0 \tag{1}$$

where $\rho$ is the mass density of magnetic fluids, $t$ is the time, and $\boldsymbol{u}$ is the velocity vector of magnetic fluids. The momentum balance equation for LMMFs reads as [27]:

$$\rho\left[\frac{\partial \boldsymbol{u}}{\partial t}+(\boldsymbol{u}\cdot\nabla)\boldsymbol{u}\right]=-\nabla p+\rho \boldsymbol{f}+(\eta+\eta_2)\nabla(\nabla\cdot\boldsymbol{u})+\eta\nabla^2\boldsymbol{u}+\nabla\times[2\eta_1(\boldsymbol{\omega}_p-\nabla\times\boldsymbol{u})] \\ \rho_e\boldsymbol{E}+\boldsymbol{j}\times\boldsymbol{B}+(\boldsymbol{P}\cdot\nabla)\boldsymbol{E}+(\boldsymbol{M}\cdot\nabla)\boldsymbol{H}+\boldsymbol{P}\times(\nabla\times\boldsymbol{E})+\boldsymbol{M}\times(\nabla\times\boldsymbol{H}) \tag{2}$$

where $p$ is the pressure of magnetic fluids, $\boldsymbol{f}$ the body force per unit mass, $\eta$ the first coefficient of viscosity (shear viscosity), $\eta_2$ the second coefficient of viscosity, $\eta_1$ the vortex viscosity, $\boldsymbol{\omega}_p$ the averaged rotational angular velocity of magnetic particles, $\rho_e$ electric charge density, $\boldsymbol{j}$ the electric current density vector, $\boldsymbol{B}$ the magnetic flux density vector, $\boldsymbol{P}$ the volumetric polarization vector, $\boldsymbol{E}$ the electric field strength vector, $\boldsymbol{M}$ the volumetric magnetization vector, and $\boldsymbol{H}$ the magnetic field strength vector. Electromagnetic fields obey Maxwell's equations:

$$\nabla\cdot\boldsymbol{D}=\rho_e \tag{3}$$

$$\nabla\times\boldsymbol{E}=-\frac{\partial \boldsymbol{B}}{\partial t} \tag{4}$$

$$\nabla\cdot\boldsymbol{B}=0 \tag{5}$$

$$\nabla\times\boldsymbol{H}=\frac{\partial \boldsymbol{D}}{\partial t}+\boldsymbol{j} \tag{6}$$

where $\boldsymbol{D}$ is the electric displacement vector. $\boldsymbol{D}$ and $\boldsymbol{B}$ are given by:

$$\boldsymbol{D}=\varepsilon_0\boldsymbol{E}+\boldsymbol{P} \tag{7}$$

$$\boldsymbol{B}=\mu_0\boldsymbol{H}+\boldsymbol{M} \tag{8}$$

where $\varepsilon_0$ is the dielectric constant of vacuum, and $\mu_0$ is the permeability of free space. $\boldsymbol{j}$ is given by the generalized Ohm's law:

$$\boldsymbol{j}=\sigma_e(\boldsymbol{E}+\boldsymbol{u}\times\boldsymbol{B})+\rho_e\boldsymbol{u} \tag{9}$$

where $\sigma_e$ is the electrical conductivity of LMMFs.

Spin-diffusion theory is widely used to explain the motion of magnetic fluids in rotating magnetic fields [3]. The spin-diffusion theory posits that the rotating magnetic field induces rotation of magnetic particles within the fluids. The angular momentum of these particles is transferred to the carrier liquids via their surfaces, thereby generating the overall angular momentum of the magnetic fluid. For LMMFs, the effect of the rotating magnetic field on the carrier liquid itself cannot be ignored (primarily due to the $\boldsymbol{j}\times\boldsymbol{B}$ term in the equation). The forces acting on the pure LM droplet can be decomposed into a driving force on the mass-center of the droplet and a torque about that center. Since the droplet is a liquid, the torque causes flow in the slip layer at its surface, but this torque contributes almost nothing to the motion of the mass-center of the droplet [15]. The relative magnitude of two effects, ferrohydrodynamic and magnetohydrodynamic, is highly dependent on the magnetic field strength. At low magnetic fields, the motion of magnetic particles dominates. Each rotating particle drives the rotation of the surrounding LM carrier liquid. The vorticity generated by multiple rotating particles combines, synthesizing the overall vorticity of the LM. This scenario closely resembles the behavior of traditional magnetic fluids under a rotating magnetic field. When the magnetic field is strong, the effect of the varying magnetic field on LM becomes significant.

**3.2. Theory of Droplet Separation and Aggregation**

Energy balance analysis is applied to the phenomena of droplet stretching, separation, and aggregation [28]. The droplet collision process is divided into three stages. Stage 1 spans from droplet impact until the collision normal velocity reaches zero; Stage 2 is the reflexive or stretching separation stage; Stage 3 is the rotational separation stage driven by non-zero angular momentum resulting from the collision. In the following analysis, all physical quantities are normalized by $8\pi R^2\sigma$, where $R$ is the droplet radius, and $\sigma$ is the surface tension coefficient. In Stage 1, viscous losses during collision primarily originate from the stagnation region. The dimensionless form of viscous loss, $\Phi_1$, is [28]:

$$\Phi_1 = \alpha \mathrm{We}\left(1 - X^2\right) \tag{10}$$

where the proportionality coefficient $\alpha$ is determined by experimental fitting, $X$ is the impact parameter, and We is the collision Weber number given by:

$$\mathrm{We} = \frac{\rho D_l U^2}{\sigma} \tag{11}$$

where $D_l$ is the droplet diameter, and $U$ the relative velocity of the two collision droplets. During Stage 2, viscous dissipation originates from the reflexive flow $\Phi_{2,r}$, the stretching flow $\Phi_{2,s}$, and the spin-induced internal rotating flow $\Phi_{2,spin}$. Reflexive flow is driven by surface tension, satisfying the relation $\rho U_{n,r}^2/2 \sim \sigma/R$; stretching flow can be approximated as $U_s = UX$. The characteristic time for reflexive separation is estimated to be on the order of the natural oscillation period $\tau = (\rho R^3/\sigma)^{1/2}$, while the characteristic time for stretching separation is approximately $2R/U_s$. Thus, viscous dissipation of reflexive flow is estimated to be $\Phi_{2,r} \sim \mu(U_{n,r}/R)^2(4\pi R^3/3)(\rho R^3/\sigma)^{1/2}/(8\pi R^2\sigma)$, and viscous dissipation of stretching flow is estimated to be $\Phi_{2,s} \sim \mu(U_s/R)^2(4\pi R^3/3)(2R/U_s)/(8\pi R^2\sigma)$. The droplet spin induces a linear velocity $U_r \sim R\omega_d$, leading to additional viscous dissipation $\Phi_{2,spin} \sim \mu(U_r/R)^2(4\pi R^3/3)(\rho R^3/\sigma)^{1/2}/(8\pi R^2\sigma)$. The total viscous dissipation $\Phi_2$ is estimated as [28]:

$$\Phi_2 = \Phi_{2,r} + \Phi_{2,s} + \Phi_{2,spin} = \mathrm{Oh}\left(\beta_1 + \beta_2\sqrt{\mathrm{We}X^2} + \beta_3 \mathrm{We}_s\right) \tag{12}$$

where Oh is the droplet Ohnesorge number given by:

$$\mathrm{Oh} = \frac{\eta}{\sqrt{\rho D_l \sigma}} \tag{13}$$

$\mathrm{We}_s$ is the spin Weber number given by:

$$\mathrm{We}_s = \frac{\rho D_l (R\omega_d)^2}{\sigma} \tag{14}$$

where $\omega_d$ is the spin angular velocity of the droplet. In Stage 3, the non-zero angular momentum from the collision is converted into the rotational angular momentum of the merged droplet. The rotational kinetic energy $E_r$ is [28]:

$$E_r = \frac{\beta_4 \mathrm{We}_{\mathrm{eff}} X^2}{\left(\beta_5 \mathrm{We}_{\mathrm{eff}} X^2\right)^{-2} + \left(\beta_5 \mathrm{We}_{\mathrm{eff}} X^2\right)^2} \tag{15}$$

where $\mathrm{We}_{\mathrm{eff}} X^2 = [(\mathrm{We}X^2)^{1/2} + \beta_6 \mathrm{We}_s^{1/2}]^2$ is the effective tangential component of the stretching inertia. The coefficients $\beta_1$, $\beta_2$, $\beta_3$, $\beta_4$, $\beta_5$, and $\beta_6$ in the above equations are experimentally determined proportional coefficients. The initial kinetic energy $E_{k0}$ before the collision can be estimated as We. If the initial droplet is spinning, an additional term must be included: $E_{k0,s} \sim [(2/5)(4\pi R^3\rho/3)R^2\omega_d]^2/2[2(4\pi R^3\rho/3)R^2/5](8\pi R^2\sigma) \sim \mathrm{We}_s/60$. The separation criterion for droplet stretching can be expressed as: The sum of initial kinetic energy ($E_{k0}$) and initial surface energy ($E_{s0}$) must exceed the sum of viscous dissipation, rotational kinetic energy of merged droplets ($E_r$), and surface energy of droplets at the point of separation ($E_s$). The formula can be expressed as [28]:

$$E_{k0} + E_{s0} > \Phi_1 + \Phi_2 + E_r + E_s \tag{16}$$

The dimensionless form of the initial kinetic energy is We. Substituting all the above expressions yields:

$$\text{We} + \frac{\text{We}_s}{60} > \alpha \text{We}(1 - X^2) + \beta_1 \text{Oh}\left(1 + \frac{\beta_2}{\beta_1}\sqrt{\text{We}X^2} + \frac{\beta_3}{\beta_1}\text{We}_s\right)$$
$$+ \frac{\beta_4 \text{We}_{\text{eff}} X^2}{\left(\beta_5 \text{We}_{\text{eff}} X^2\right)^{-2} + \left(\beta_5 \text{We}_{\text{eff}} X^2\right)^2} + E_s - E_{s0} \tag{17}$$

Consider the simplest case where the droplets collide head-on and their relative velocity is negligible compared to the droplet spin angular velocity, so we can get $X = 0$ and $U = 0$. Therefore, we can simplify the separation criterion to:

$$\text{We}_s > \frac{60(\beta_1 \text{Oh} + E_s - E_{s0})}{1 - 60\text{Oh}\beta_3} \tag{18}$$

The right term of the criterion is independent of the LMMF droplet spin angular velocity $\omega_d$. Therefore, it can be inferred that when $\omega_d$ increases beyond the limit of its surface energy change, the droplet will separate. The droplet spin angular velocity $\omega_d$ is strongly correlated with both the magnetic field rotation speed and magnetic field strength, necessitating a comprehensive determination through both experimental and theoretical approaches.

## 4. Typical Fluid Patterns Disclosed from Experiments

In this section, we present typical LMMF patterns and classifications in the Taiji analogous pattern and phase diagram. These patterns are closely related to magnetic field strength, magnetic field rotational speed, and droplet volume, respectively.

### 4.1. Single Droplet Spin and Deformation

Before magnetic field rotation begins, the LMMF droplet rests off-center on one magnet, attracted by the maximum field gradient present at a single pole. Once high-frequency rotation starts, the periodic variation of the axial magnetic flux density gradient creates axially symmetric forces, confining the droplet at the center of the magnet pair. This magnetic trapping is mechanistically similar to the behavior of Tesla's Egg of Columbus or a magnetic stir bar. We quantify the minimum rotation speed, $\omega_c$, required for this center confining effect as presented in Figure 5(a). The results indicate that the required $\omega_c$ increases with magnetic field strength. This result is reasonable. As shown in Figure 1(f), at low distances where the magnetic field is stronger, two magnetic field gradient peaks exist in the $B_x$ formed by the magnetic pair. This necessitates a higher magnetic field rotation speed, causing the magnetic field change to significantly exceed the magnetic relaxation time of the magnetic particles. Only then can the LMMF droplet overcome the circumferential magnetic gradient inhomogeneity.

Figure 5(b) illustrates the circular motion of an LMMF droplet around a magnet (see also Movie S1). This phenomenon typically occurs under conditions of strong magnetic fields and low field rotational speeds ($\omega < \omega_c$). It can be observed that due to the strong magnetic field, the droplet is flattened with its shape resembling a flattened ellipse. The period of the droplet circular motion also aligns with the period of the magnet movement.

For weak magnetic fields at low rotational speeds ($\omega > \omega_c$), the LMMF droplet spins at the center in response to the rotating magnetic field. In our experiment, the LMMF droplet assumes an ellipsoidal shape at low field rotational speeds, completing one rotation cycle as depicted in Figure 5(c). As $\omega$ increases, the LMMF droplet gradually elongates into a dumbbell shape, wider at both ends and narrower in the middle, with one rotation cycle shown in Figure 5(d). Obviously, the droplet undergoes significant shape and spin angular velocity changes with field rotational speed. This result differs significantly from classical non-conducting ferrofluids. Their droplets also exhibit an ellipsoidal shape under weaker magnetic fields. However, as the magnetic field rotational speed increases, classical ferrofluid droplets maintain their ellipsoidal shape without undergoing significant morphological transformation [5][6]. Taking a 300 μL LMMF droplet as an example, we present experimental plots of droplet angular velocity ($\omega_d$) and droplet shape parameter ($a/c$ and $a$) versus magnetic field rotation speed in Figures 5(e) and 5(f), respectively. Before undergoing shape transformation, $\omega_d$ is nearly identical to $\omega$. However, after transforming into a dumbbell shape, $\omega_d$ gradually decreases as the magnetic field speed increases, and the LMMF droplet undergoes asynchronous spinning (see Movie S2). We attribute this phenomenon to magnetic particle relaxation. When the magnetic field period becomes shorter than the magnetic relaxation time, the rotation of magnetic particles no longer synchronizes with the magnetic field rotation. During asynchronous rotation, the direction of the LMMF magnetization (***M***) exhibits a phase difference relative to the magnetic field direction (***H***). Consequently, the ***M*** can be decomposed into two components: one parallel to the ***H*** and another perpendicular to ***H***. The droplet tends to elongate along the direction of the stronger magnetization component, thereby minimizing the total energy.

    For stronger magnetic fields and larger droplet volumes, the LMMF droplet splits into smaller droplets, each continuing its own rotation.

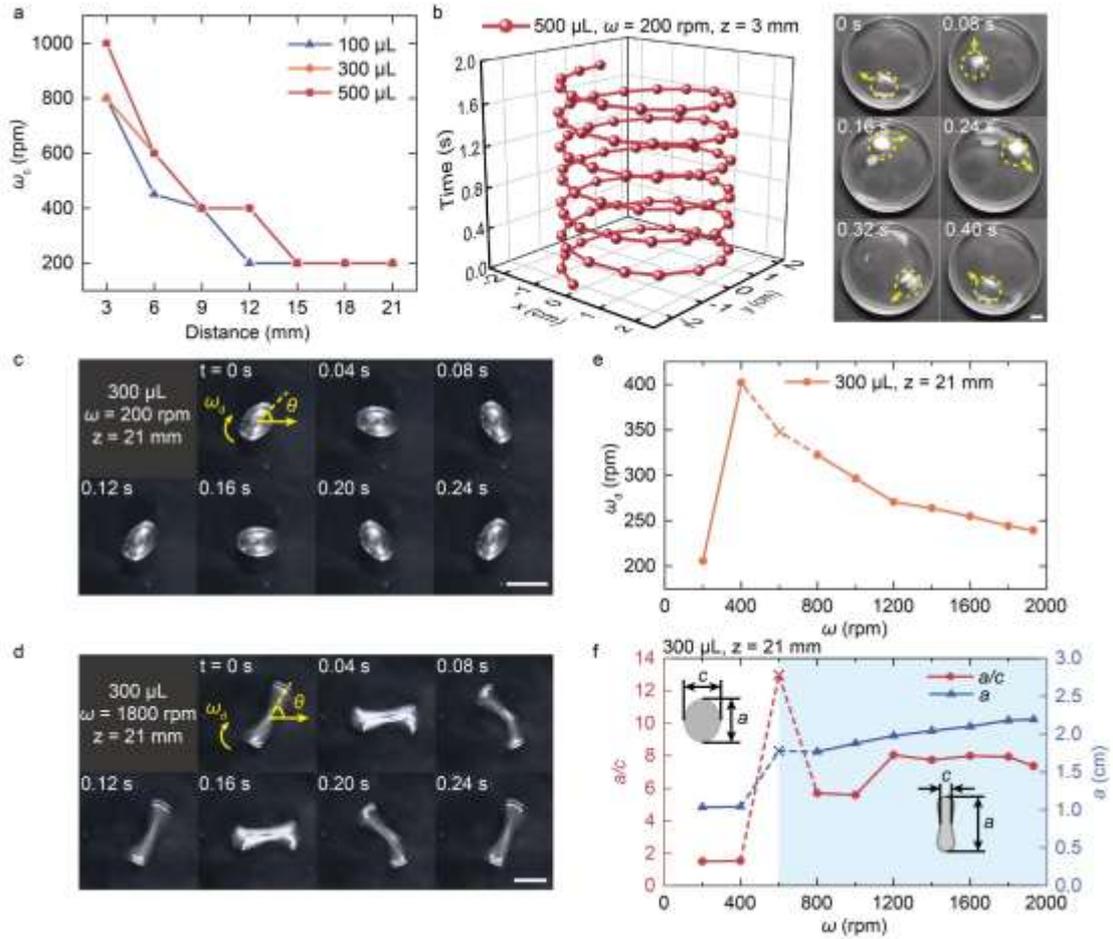

Figure 5. Single LMMF droplet motion patterns and characterization. (a) The minimum magnetic field rotational speed, $\omega_c$, required to center droplets of different volumes under varying magnetic field strengths. (b) Time-displacement curve of a 500 μL droplet moving with a magnet at low rotational speed, along with a set of photographs corresponding to one full rotation (z = 3 mm, $\omega$ = 200 rpm). (c) Schematic of an ellipsoidal droplet completing one full rotation (z = 21 mm, $\omega$ = 200 rpm). (d) Schematic of a dumbbell droplet completing one full rotation (z = 21 mm, $\omega$ = 1800 rpm). (e) The spin angular velocity of 300 μL droplet as a function of magnetic field rotation speed (z = 21 mm). The cross mark indicates where the droplet separates at that location, with the value representing the spin angular velocity prior to separation. (f) The aspect ratio of droplets as a function of magnetic field rotational speed. The cross mark indicates where the droplet separates at that location, with the values representing the aspect ratio and long axis length before separation. The blue region indicates where droplets assume a dumbbell shape. The scale bars are all 1 cm.

**4.2. Droplet Separation**

As mentioned in Section 2.2, LMMF droplets can only maintain a single state under low magnetic field strength and field rotational speed. As droplet volume increases and magnetic field strength intensifies, LMMF droplets tend to split into multiple smaller droplets. Figure 6(a) shows the minimum magnetic field rotational speed required for droplet separation ($\omega_s$) under different conditions. The results indicate that larger droplets require a lower $\omega_s$ for separation, meaning they are easier to split. Additionally, weaker magnetic fields result in greater $\omega_s$, making droplet separation more difficult. Figure 6(b) illustrates a typical droplet separation and coalescence process (see also Movie S3). During high-speed rotation, LMMF droplets initially form a dumbbell shape, with two spherical ends connected by a long ligament. Due to excessive rotational speed, surface tension struggles to maintain a single droplet, causing the connecting ligament to break and the droplet to split into two. Under the influence of the rotating magnetic field, the separated droplets each begin to spin and develop into dumbbell shapes. However, the smaller droplets, being smaller in volume, require less surface tension to maintain their shape and thus do not split further. After a brief separation, due to the tendency toward centripetal motion, the split droplets move toward the center while spinning. Consequently, at a certain point, the two droplets collide and merge. This separation and coalescence process will continue to repeat cyclically.

Due to the small volume of 100 μL droplets, they disperse more widely after separation and do not coalesce quickly. This makes it easy to capture multiple dispersed droplets through imaging, facilitating analysis of the relationship between droplet separation and magnetic field strength and field rotational speed. Figure 6(c) shows the relationship between the number of droplets after separation and $\omega$ under varying magnetic field strength. Under strong magnetic fields and high field rotational speeds, LMMF droplets are likely to be dispersed into a very large number of small droplets. Figure 6(d) shows grayscale images of numerous droplets separated at $\omega$ from 1000 rpm to 1930 rpm. From an energy perspective, strong magnetic fields and high rotational speeds impart significant rotational kinetic energy to LMMF droplets, causing them to tend toward separation to reduce energy. Simultaneously, the resulting separated droplets undergo centripetal collisions, further intensifying this separation process.

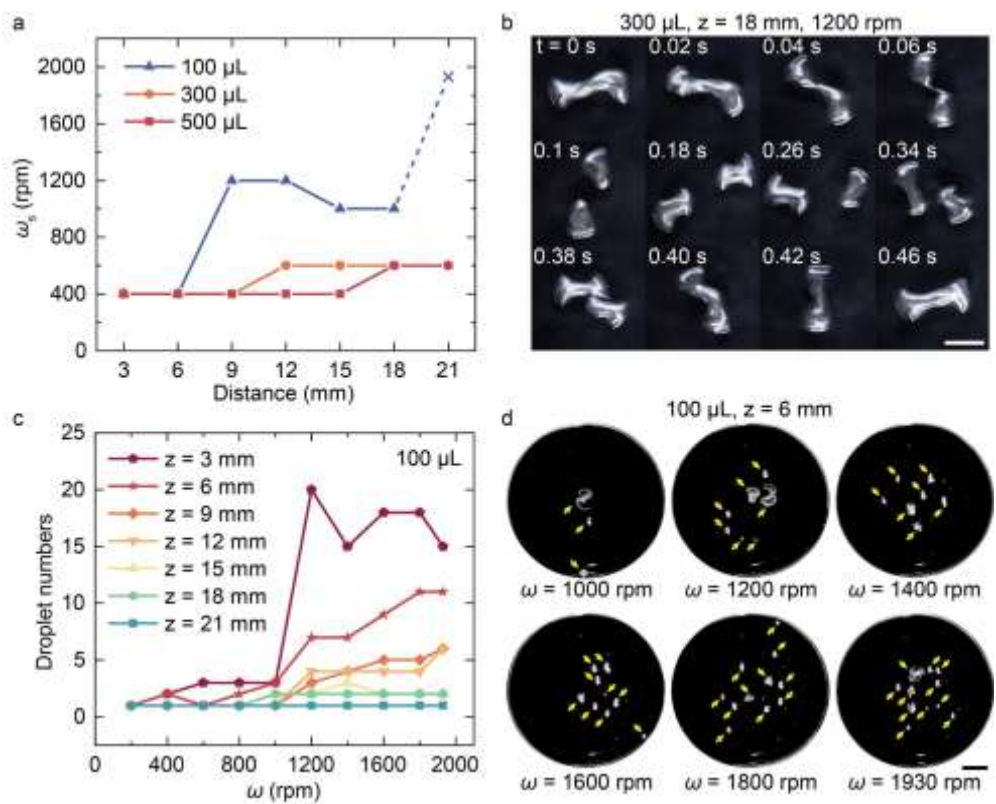

Figure 6. Droplet separation characteristics. (a) The minimum magnetic field rotation speed required for droplets of different volumes to split under varying magnetic fields. (b) Schematic of 300 μL droplet separation and coalescence within a cycle (z = 18 mm, $\omega$ = 1200 rpm). (c) The number of splitting droplets as a function of magnetic field rotational speed under different magnetic field

strengths. This number refers to the maximum number of splitting droplets occurring during the experimental filming process. (d) A set of grayscale images corresponding to photographs of droplets splitting into multiple droplets under certain conditions. Yellow arrowheads mark each split droplet. The scale bars are all 1 cm.

Under conditions of relatively high magnetic field strength and low magnetic field rotational speed, LMMF droplets are found to exhibit a multi-orbit circular motion (Movie S4). For a 100 μL LMMF droplet, at z = 3 mm and $\omega$ = 400 rpm, the droplet splits into two (Figure 7(a)). One droplet moves in a circular path within the inner ring at a rotational speed equivalent to the magnetic field rotational speed (~ 400 rpm), while the other droplet moves in a slow circular path at the outermost ring at a rotational speed of 25.7 rpm. As $\omega$ increases, the droplet splits into a greater number, revealing three concentric orbits of differing radii in motion (Figure 7(b)). At $\omega$ = 800 rpm, the outer-most LMMF track rotates at approximately 68.3 rpm, while the middle LMMF track rotates at a speed comparable to that of the magnet (~800 rpm). The inner-most track rotates at approximately 73.6 rpm. When the magnetic field is slightly weakened (z = 6 mm), more small droplets separate and move closer to the inner ring (Figure 7(c)). For larger LMMF droplets, a similar multi-orbit motion phenomenon occurs, but it stirs the hydrochloric acid solution more vigorously. This intense agitation increases the shear force between the LMMF and the solution, causing the LMMF to fragment more completely (Figure 7(d)-7(g)). Specifically, under moderate magnetic fields and low rotational speeds, galaxy-like stellar and satellite motions may emerge due to differences in volume or magnetic particle content. Figure 7(h) illustrates one typical satellite motion (see also Movie S5). After droplet separation, a smaller droplet is captured at the center and undergoes rotation, while the larger droplet follows a nearly square orbital motion as it rotates with the magnetic field. The orbital curve of the satellite droplet is shown in Figure 7(i). The major axis of the orbit has also been found to be deflecting over time, deviating by approximately 10° per cycle.

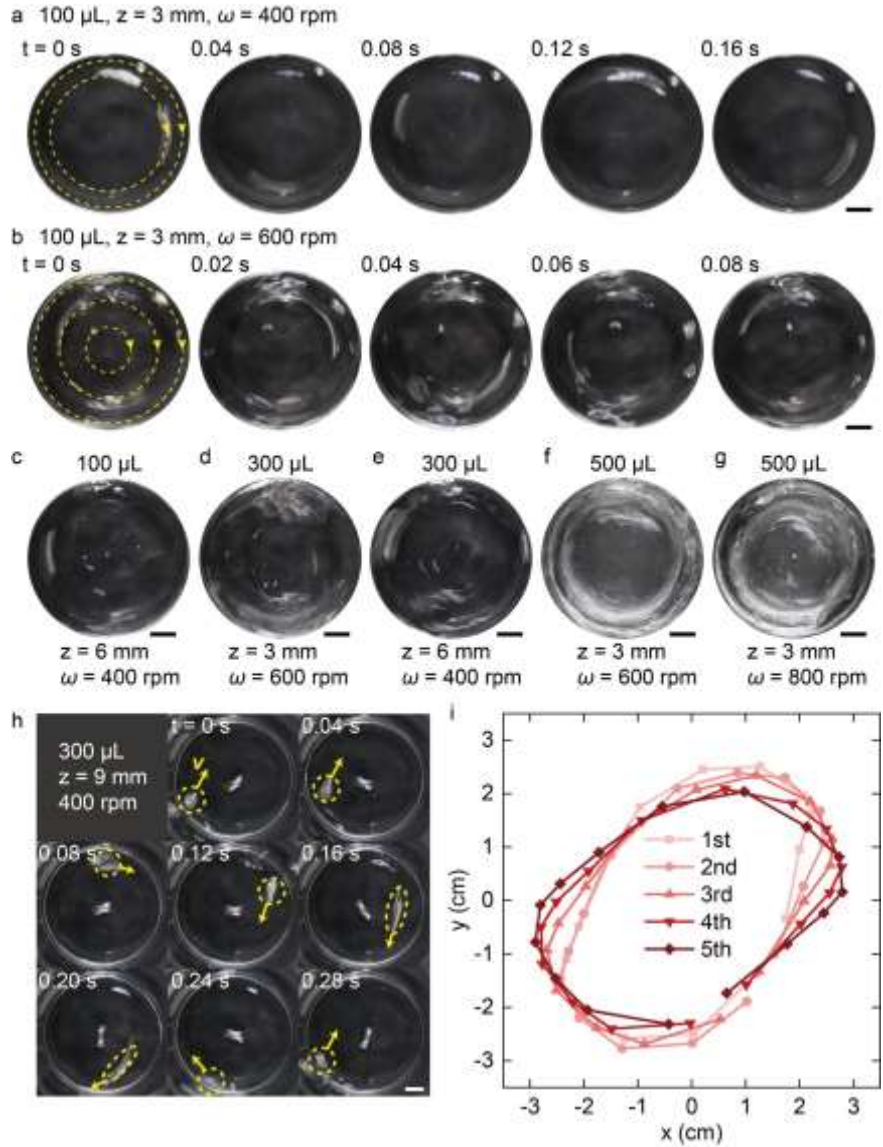

Figure 7. Multi-orbital motion of LMMF droplets. (a) Flow pattern of 100 μL LMMF droplet within 0.16 s ($z = 3$ mm, $\omega = 400$ rpm). The yellow dashed lines represent the trajectory and direction of droplet movement. (b) Flow pattern of 100 μL LMMF droplet within 0.08 s ($z = 3$ mm, $\omega = 600$ rpm). The yellow dashed lines represent the trajectory and direction of droplet movement. (c) Flow pattern of 100 μL LMMF droplet ($z = 6$ mm, $\omega = 400$ rpm). (d) Flow pattern of 300 μL LMMF droplet ($z = 3$ mm, $\omega = 600$ rpm). (e) Flow pattern of 300 μL LMMF droplet ($z = 6$ mm, $\omega = 400$ rpm). (f) Flow pattern of 500 μL LMMF droplet ($z = 3$ mm, $\omega = 600$ rpm). (g) Flow pattern of 500 μL LMMF droplet ($z = 3$ mm, $\omega = 800$ rpm). (h) Two separated droplets undergo distinct periodic motions. One droplet is captured at the center and rotates on its own axis, while the other satellite droplet performs a square-shaped motion around the central droplet. (i) The displacement curve diagram of the satellite droplet depicts the behavior over five consecutive cycles. The movement trajectories are all clockwise. The scale bars are all 1 cm.

For large droplet volumes (300 μL or 500 μL), the dispersion effect is diminished at high field rotational speed, and LMMF droplets may exhibit peculiar separation phenomena. At low magnetic field strengths, the LMMF droplet may eject smaller droplets when the magnetic field rotational speed increases (Figure 8(a)). Based on the droplet motion, it is inferred that the ejected droplet contains very low magnetic particle content, which is unlikely to return to the original droplet. At moderate magnetic field strengths, droplet separation also occurs as the magnetic field rotational speed increases, but the separation process differs significantly from that observed at low magnetic field strengths. Taking a 500 μL droplet as an example, spinning it at $\omega$ = 800 rpm and z = 9 mm will fling out numerous small droplets (Figure 8(b) and Movie S6). Some of these small droplets will strike the container edge, causing them to orbit around the edge rather than return to the center of the magnetic pair. The small droplets are observed to display fluid motion comparable to pure LM, suggesting their reduced magnetic particle content. When the magnetic field rotational speed is increased to 1000 rpm, the ejected droplets become more susceptible to centrifugal force and return toward the center of the magnet pair (Figure 8(c) and Movie S6). This may be attributed to that the increased $\omega$ reduces the droplet spin angular velocity, thereby decreasing the centrifugal force acting on the droplet, as shown in Figure 5(e).

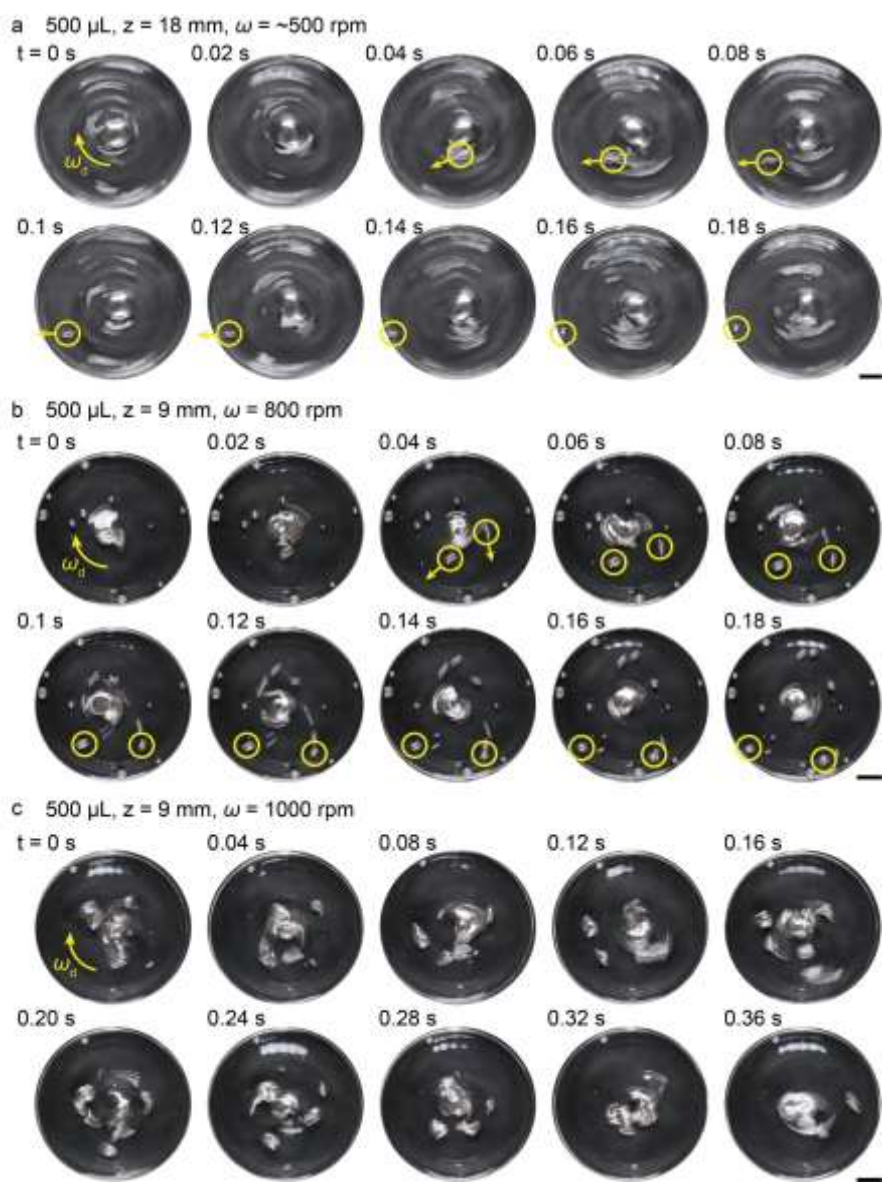

Figure 8. Peculiar separation phenomenon for large-volume LMMF droplets. (a) The process of ejecting one small droplet from a spinning 500 μL LMMF droplet (z = 18 mm, $\omega \approx$ 800 rpm). Yellow

circles indicate the motion of the ejected droplet. (b) The process of ejecting a large number of small droplets from a spinning 500 μL LMMF droplet (z = 9 mm, $\omega$ = 800 rpm). Yellow circles indicate the motion of two typical ejected droplets. (c) The separation and coalescence process of 500 μL LMMF droplets (z = 9 mm, $\omega$ = 1000 rpm). The scale bars are all 1 cm.

### 4.3. Torus Flow Patterns

Under specific conditions, LMMF droplets can form a stable torus flow as shown in Figure 9(a) and Movie S1. Unlike previous studies, this torus flow requires no solid wall confinement and does not dissipate over time, instead persisting in a stable manner [29]. The two peaks of $B_x$ shown in Figure 1(f) replace the function of the solid wall confinement. We attribute the formation of the torus flow to a state of equilibrium achieved through the centripetal motion of magnetic particles and the centrifugal motion of the LM carrier liquid. To validate our hypothesis, we demonstrated different flow patterns under conditions of higher or lower $\omega$. Figure 9(b) shows that as $\omega$ increases, $\omega_d$ decreases and centripetal force becomes dominant, causing the torus to shrink. Conversely, Figure 9(c) indicates that as $\omega$ decreases, $\omega_d$ increases and centrifugal force becomes dominant, causing the torus to expand. In both cases, it is difficult to form a torus. On the other hand, if the magnetic field strength is reduced while $\omega$ remains unchanged, the annular structure is also difficult to form (Figure 9(d)). For greater distances z, the toroidal wave peaks formed by $B_x$ are significantly attenuated, insufficient to generate centrifugal forces strong enough to create central holes in LMMF droplets. When the LMMF droplet volume is reduced, the separated droplets are found to move in a quasi-ring pattern, but due to their small size, they struggle to form a complete torus. Under the same conditions that formed a ring with a 500 μL droplet, a 300 μL droplet is observed to separate into multiple droplets performing quasi-ring motion (Figure 9(e)). When $\omega$ is reduced to 800 rpm, this quasi-ring motion becomes more pronounced (Figure 9(f) and Movie S7).

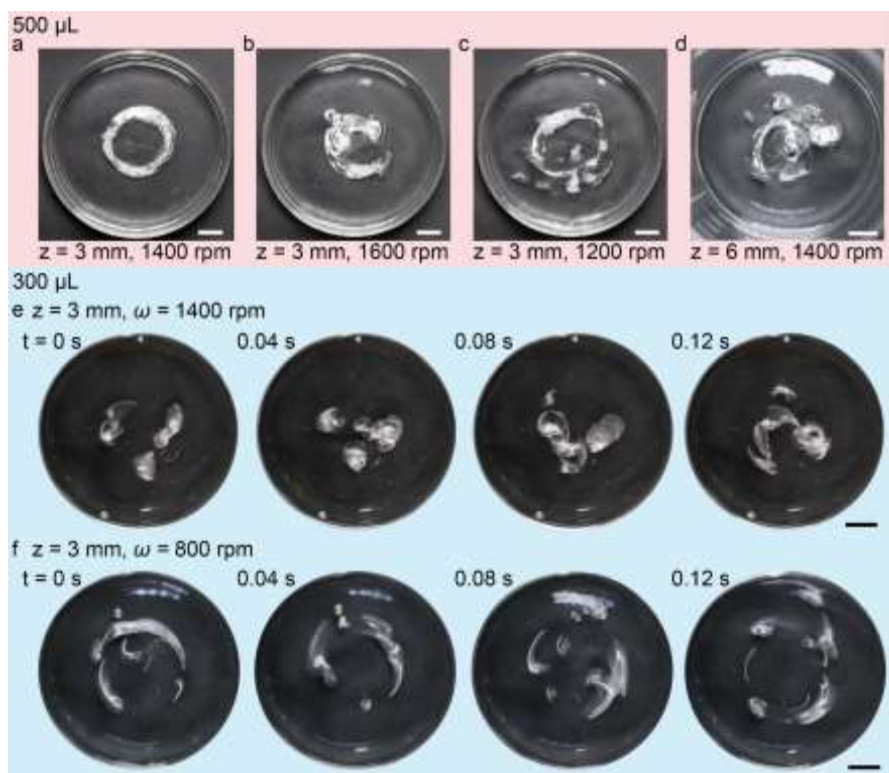

Figure 9. Torus flow characteristics. (a) The stable torus flow pattern formed by a 500 μL droplet (z

= 3 mm, $\omega$ = 1400 rpm). (b) The unstable flow pattern formed by a 500 μL droplet (z = 3 mm, $\omega$ = 1600 rpm). (c) The unstable flow pattern formed by a 500 μL droplet (z = 3 mm, $\omega$ = 1200 rpm). (d) The unstable flow pattern formed by a 500 μL droplet (z = 6 mm, $\omega$ = 1400 rpm). (e) The unstable quasi-ring pattern formed by a 300 μL droplet over a period of time (z = 3 mm, $\omega$ = 1400 rpm). (f) The unstable quasi-ring pattern formed by a 300 μL droplet over a period of time (z = 3 mm, $\omega$ = 800 rpm). The scale bars are all 1 cm.

### 4.4. Taiji-like Flow Patterns

Under specific conditions (z = 3 mm, $\omega$ = 1930 rpm), 500 μL LMMF droplets form peculiar patterns resembling the interlocking black and white shapes in the Taiji symbol or the collision of two galaxies, as shown in Figure 10(a) and Movie S1. This may occur because the trailing flow tends toward annular motion while the leading flow tends toward centripetal spin, balancing to form this hook-shaped droplet. The observed Taiji-like flow pattern can be viewed as a vivid embodiment of the Taiji philosophy in LMMF dynamics. The rotating magnetic field and magnetic particles form the two poles of the system, i.e., Yin and Yang. The periodic separation of the LMMF, akin to "when Yang reaches its peak, Yin emerges," represents the transformation of "Yang giving birth to Yin" after the energy of the system (magnetic energy, surface energy, kinetic energy) accumulates to a critical point. The subsequent coalesce corresponds to "Yin giving birth to Yang." This embodies the dynamic process of "Yin-Yang waxing and waning," where different energies continuously transform and balance each other through mutual conversion.

When $\omega$ is reduced, the pattern of Yin-Yang separation periodically weakens. During this transition, three hook-shaped droplets may appear transiently (Figure 10(b)). This may be due to the slightly increased centrifugal force at lower $\omega$, which makes droplet separation more pronounced. Further reducing the magnetic field strength weakens this separation effect. The pattern transforms into a high-speed spin of the central droplet, accompanied by the elongation and distortion of the long snakelike LMMF as it orbits around the central droplet (Figure 10(c) and Movie S8). This is because, as the distance increases, the toroidal potential barrier weakens, causing the LMMF droplet to preferentially spin at the center of the magnet pair. For smaller droplets (300 μL), separation becomes asymmetric, and the number of separated droplets increases, which is consistent with the analysis in section 2.2 (Figure 10(d)).

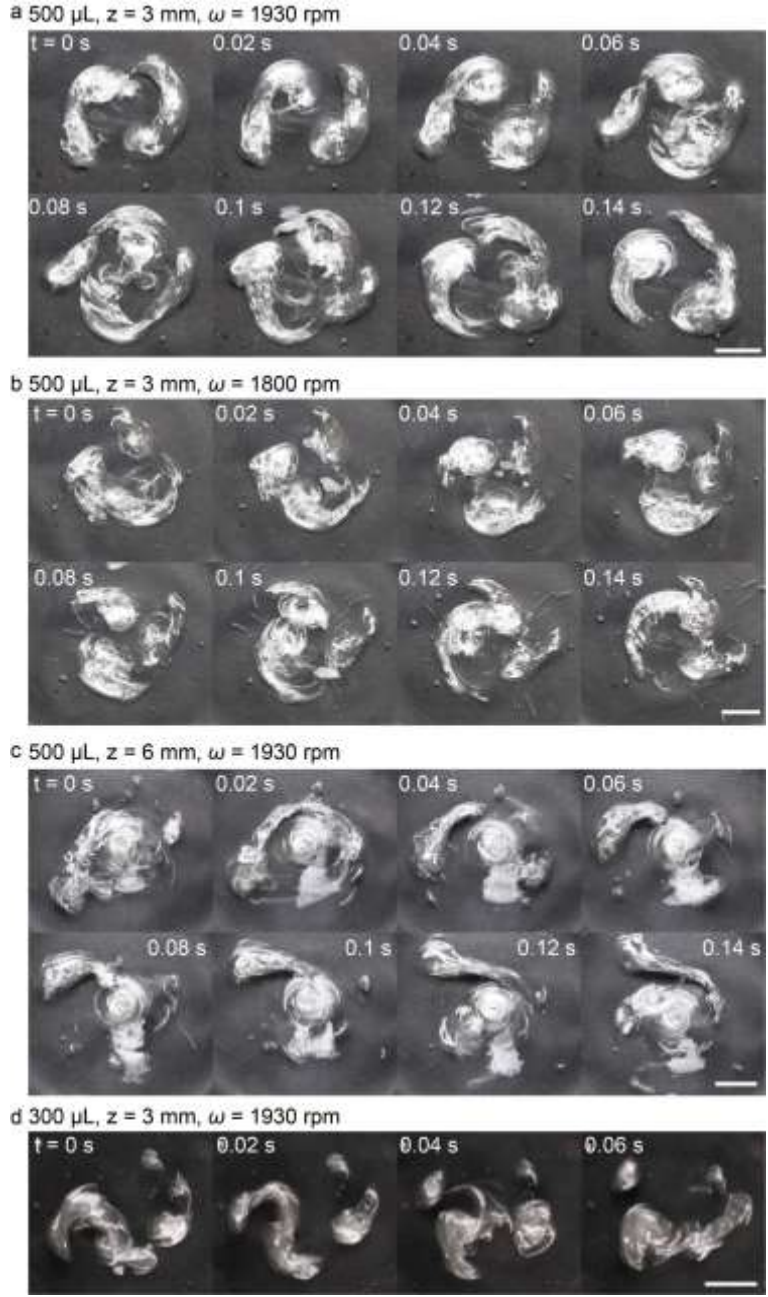

Figure 10. Taiji-like flowing pattern analysis. (a) Taiji-like flowing of 500 μL LMMF droplet within a cycle (z = 3 mm, $\omega$ = 1930 rpm). (b) Flow pattern of 500 μL LMMF droplet within 0.14 s (z = 3 mm, $\omega$ = 1800 rpm). (c) Flow pattern of 500 μL LMMF droplet within 0.14 s (z = 6 mm, $\omega$ = 1800 rpm). (d) Flow pattern of 300 μL LMMF droplet within 0.06 s (z = 3 mm, $\omega$ = 1930 rpm). The scale bars are all 1 cm.

Under moderate magnetic fields, LMMF droplets exhibit pronounced twisting deformation when exposed to high-speed rotating magnetic fields. For a 500 μL droplet, it twists clockwise toward the center of the magnet pair (Figure 11(a)). The wake of the droplet extends outward but also spins clockwise with the droplet. Upon reaching a certain length, the droplet breaks and contracts, then repeats the twisting and elongation cycle. For a 300 μL droplet, twisting also occurs, though the elongation length is significantly shorter than that of 500 μL droplets (Figure 11(b)). This deformation mechanism is analogous to the force mechanism described in Section 4.1, where deformation results from the combined action of a normal force parallel to ***H*** and a tangential force perpendicular to ***H***. The difference lies in the fact that under moderate magnetic fields, the normal force controlling droplet elongation increases beyond the limit that surface tension can sustain. Consequently, the droplet shape does not stabilize into a dumbbell configuration but instead undergoes continuous stretching. Simultaneously, the tangential force induces rotation in the elongated droplet until it stretches to separation. Under these magnetic field conditions, the shape of the LMMF droplet remains in a state of dynamic equilibrium, making it difficult to maintain stability.

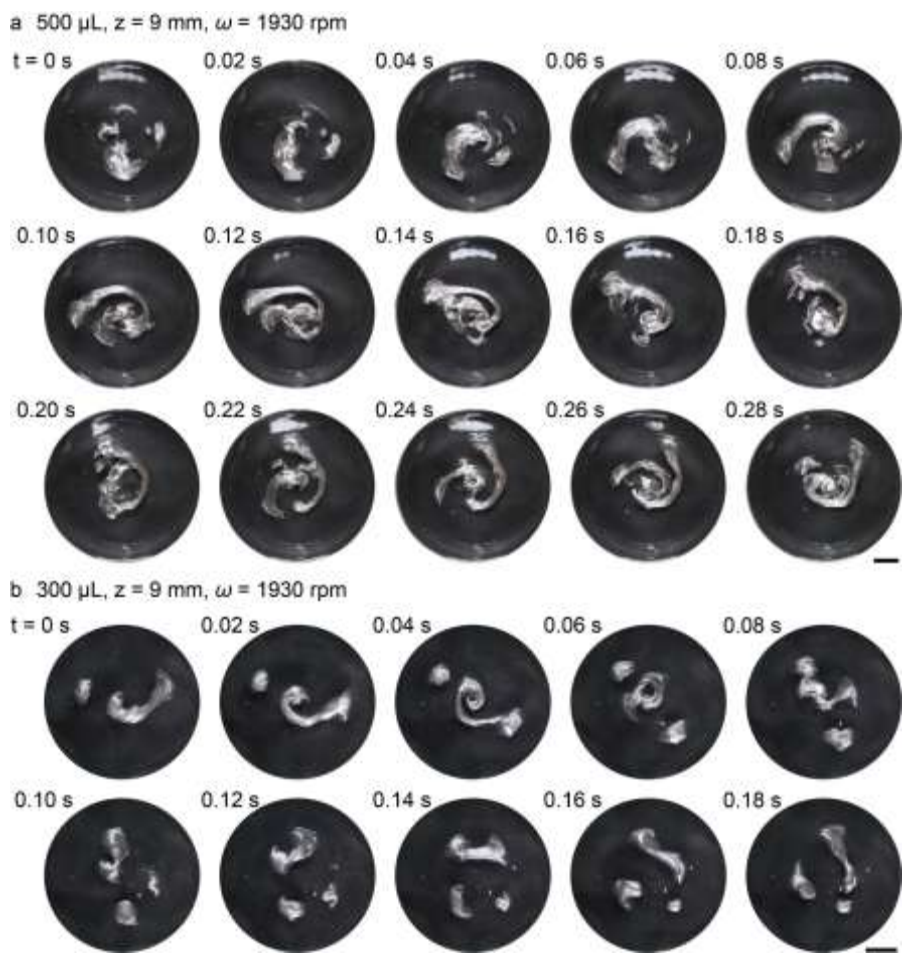

Figure 11. Spin and twisting effect of LMMF droplet. (a) Flow pattern of 500 μL LMMF droplet within 0.28 s ($z$ = 9 mm, $\omega$ = 1930 rpm). (b) Flow pattern of 300 μL LMMF droplet within 0.18 s ($z$ = 9 mm, $\omega$ = 1930 rpm). The scale bars are all 1 cm.

## 5. Discussions and Perspectives

Conventional magnetic fluid studies typically rely on non-conducting or low-conductivity carrier liquids, with theoretical models primarily focusing on the coupling between magnetic fields and fluid motion. This work introduces electrically conductive LMs as the carrier liquid, which significantly enhances electromagnetic induction effects within the system and introduces key physical processes such as electromagnetic force body effects as well as eddy current dissipation. The observed complex structures, such as torus flows and Taiji patterns, result from the simultaneous competition and equilibrium between the magnetic energy of magnetic particles, the surface energy of LMs, the electromagnetic forces acting on the LMs, and other dissipation energy, respectively. However, due to complicated coupling effects, explanations on the phenomenon at the present stage still remain largely qualitative, and a thorough theoretical interpretation of all flow phenomena has yet to be achieved. It requires further refinement of theoretical frameworks and experimental justifications in the near future, along with detailed computational simulations. Previously, significant advancement has been made in the simulation of non-conducting magnetic fluids [30]. Further incorporations of the inductive effects of the magnetic liquid metal fluid need additional efforts.

Methodologically, the Taiji theory expands the experimental paradigm for multifunctional fluid patterning. Former approaches manipulated magnetic droplets at the microscopic scale or controlled pure LM flow at the macroscopic scale. This study demonstrates that various flow patterns can be achieved on demand and reversibly across a wide scale range from millimeters to centimeters. The diverse fluid patterns observed—such as torus flow, dumbbell shapes, and Taiji-like structures—can be classified as the fundamental eight trigrams. Each stable flow pattern is "encoded" by a specific combination of three core parameters: magnetic field strength, field rotational speed, and droplet volume, analogous to how each trigram is formed by the arrangement of three solid or dashed lines, containing essential information about the state of the system. Based on the above analysis, reaching the level of complexity symbolized by the sixty-four hexagrams requires further introduction of richer control conditions. First, the range of experimental conditions in this study can be extended. For instance, the control over droplet volume in this work is relatively prescribed. The behavior of bulk LMMFs or nanoscale LMMF droplets under rotating magnetic fields is worth exploring. Additionally, due to equipment limitations, the strength and frequency range of the magnetic fields as administrated in this experiment remain limited. It is necessary to introduce more complex equipment to tune such parameters. Second, discovering more flow patterns can be possible through introducing various experimental variables. Future research may explore coupling such as magnetohydrodynamic effect with electrical or electrochemical phenomena in LMs to generate additional deformation. In the present study, hydrochloric acid solution is selected as the medium. By altering the working environment (e.g., pH, surfactants) to modify the interfacial conditions between different liquids, novel fluid patterns would emerge. In addition, the type, particle size, and parameters of the magnetic particles used in this work are kept constant. Employing more other soft magnetic or hard magnetic particles, reducing particle size, or adjusting particle concentration mould yield distinct fluid patterns.

Notably, one thing should be pointed out is that the behaviors as identified so far—such as the formation of torus structures, the orbital motion of secondary droplets, and their periodic separation and coalescence—also exhibit a morphological resemblance to many macroscopic dynamical processes in the cosmos. These phenomena may be combined with planetary dynamics to aid for analyzing astrophysics through laboratory-scale experiments. Building on this work, future research can establish a "Tabletop Cosmology" experimental paradigm to capture the scaling laws and symmetries governing the physical processes. One possible fundamental strategy is to conduct experiments in microgravity or magnetic levitation environments to counteract the effects of gravity on LMMF droplet morphology and flow patterns.

Looking ahead, the controllable deformation and motion capabilities of such LMMFs promise broad application prospects. At the first impression, the kaleidoscopic dynamic fluidic patterns in fact provide a vivid display or man-machine interactional system which can be very useful for science, art and film design. Short-term goals include developing novel soft robotic actuators that achieve complex shape changes and motions, as well as constructing reconfigurable radio frequency devices (e.g., frequency-tunable ring antennas). Such effect also holds potential for developing liquid mixers or non-contact fluid manipulation systems for controlling programmable droplets. In the long term, this control system enables dynamic regulation of magnetic material fabrication, adaptive optical devices, and intelligent bionic surfaces, which can be utilized in more wide areas including thermal management or biomedical fields.

**6. Conclusions**

In summary, we have disclosed the kaleidoscopic variational phenomena and diverse motion effects of LMMFs under rotating magnets. Specifically, non-contact fixation and self-rotation of magnetic LM were achieved using a magnetic stirrer apparatus. Various flow patterns at the macroscopic scale are discovered over the experiments, such as ellipses, dumbbells, torus, and Yin-Yang patterns. Through quantifying the observations, we establish a relationship between magnet rotational speed and droplet spin angular velocity and elucidate the mechanism underlying this droplet motion. By tuning variables, we discovered that droplet separation behavior exhibits significant dependence on droplet volume, magnetic field strength, and field rotational speed. Through systematic parameter analysis and morphological classification, we reveal a profound structural and mechanism analogy between the formation and transformation of complex LMMF flow patterns and the core concepts of traditional Chinese Taiji philosophy, such as the interaction of Yin and Yang, the waxing and waning of energies, and the attainment of dynamic equilibrium. This study offers a straightforward approach for investigating magnetic fluids and reveals unconventional phenomena in conducting magnetic fluids, which opens a big space for exploring fundamental fluidic physics and also holds further potential applications in engineering areas like fluidic machinery, thermal management or biomedical fields.

**References**
[1]     Moskowitz R, Rosensweig RE. Nonmechanical torque-driven flow of a ferromagnetic fluid by an electromagnetic field. Appl Phys Lett. 1967;11(10):301–303.
[2]     Rosensweig RE, Popplewell J, Johnston RJ. Magnetic fluid motion in rotating field. J Magn Magn Mater. 1990;85(1–3):171–180.
[3]     Zaitsev VM, Shliomis MI. Entrainment of ferromagnetic suspension by a rotating field. J Appl Mech Tech Phys. 1972;10(5):696–700.
[4]     Shliomis MI. How a rotating magnetic field causes ferrofluid to rotate. Phys Rev Fluids. 2021;6(4):043701.
[5]     Bacri J-C, Cebers AO, Perzynski R. Behavior of a magnetic fluid microdrop in a rotating magnetic field. Phys Rev Lett. 1994;72(17):2705–2708.
[6]     Janiaud E, Elias F, Bacri JC, et al. Spinning ferrofluid microscopic droplets. Magnetohydrodynamics. 2000;36(4):300–311.
[7]     Cēbers A. Dynamics of an elongated magnetic droplet in a rotating field. Phys Rev E. 2002;66(6):061402.

[8] Kaiser R, Rosensweig RE. Study of ferromagnetic liquid. 1967. Available from: https://ntrs.nasa.gov/citations/19680004736.

[9] Shepherd PG, Popplewell J, Charles SW. A method of producing ferrofluid with gadolinium particles. J Phys D Appl Phys. 1970;3(12):430.

[10] Rosensweig RE. Refrigeration aspects of magnetic particle suspensions. Int J Refrig. 2006;29(8):1250–1258.

[11] Qiao M, Xing Z, Fu J, et al. Multiphase flow physics of room temperature liquid metals and its applications. Sci China Technol Sci. 2023; 66(6): 1483–1510.

[12] Bojarevics A, Beinerts T. Experiments on liquid metal flow induced by a rotating magnetic dipole. Magnetohydrodynamics. 2010;46(4):333–338.

[13] Zhao X, Yang L, Ding Y, et al. Noncoalescent liquid metal droplets sustained on a magnetic field-circulated liquid metal bath surface. Appl Phys Lett. 2019;115(8):1–5.

[14] Li N, Wang L, Liu J. Quantifying the hybrid flow and aperture forming of liquid metals immersed in NaOH solution under rotating magnetic field. Appl Res. 2024;3(1): e202200100.

[15] Shu J, Tang S-Y, Feng Z, et al. Unconventional locomotion of liquid metal droplets driven by magnetic fields. Soft Matter. 2018;14(35):7113–7118.

[16] Zhou Y, Li N, Zhao X, et al. Reversal tuning of liquid metal motor under rotating magnetic field. Adv Funct Mater. 2024;34(33):1–10.

[17] Yu Y, Miyako E. Alternating-magnetic-field-mediated wireless manipulations of a liquid metal for therapeutic bioengineering. iScience. 2018;3:134–148.

[18] Ye J, Qin P, Xing Z, et al. Liquid metal hydraulic actuation and thermal management based on rotating permanent magnets driven centrifugal pump. Int Commun Heat Mass Transf. 2022;139:106472.

[19] Xiang W, Lu Y, Wang H, et al. Liquid-metal-based magnetic fluids. Nat Rev Mater. 2024;9(6):433–449.

[20] Xiong M, Gao Y, Liu J. Fabrication of magnetic nano liquid metal fluid through loading of Ni nanoparticles into gallium or its alloy. J Magn Magn Mater. 2014;354:279–283.

[21] Carle F, Bai K, Casara J, et al. Development of magnetic liquid metal suspensions for magnetohydrodynamics. Phys Rev Fluids. 2017;2(1):013301.

[22] Sun X, Guo R, Yuan B, et al. Stiffness tunable implanted electrode enabled by magnetic liquid metal for wireless hyperthermia. Appl Mater Today. 2022;27(January):101495.

[23] Bai K, Casara J, Nair-Kanneganti A, et al. Effective magnetic susceptibility of suspensions of ferromagnetic particles. J Appl Phys. 2018;124(12):123901.

[24] Cheeseman S, Elbourne A, Kariuki R, et al. Broad-spectrum treatment of bacterial biofilms using magneto-responsive liquid metal particles. J Mater Chem B. 2020;8(47):10776–10787.

[25] Elbourne A, Cheeseman S, Atkin P, et al. Antibacterial liquid metals: Biofilm treatment via magnetic activation. ACS Nano. 2020;14(1):802–817.

[26] Tang J, Zhao X, Li J, et al. Liquid metal phagocytosis: Intermetallic wetting induced particle internalization. Adv Sci. 2017;4(5):1700024.

[27] Shizawa K, Tanahashi T. New constitutive equations for conducting magnetic fluids with internal rotation: Thermodynamical discussions. Bull JSME. 1986;29(255):2878–2884.

[28] He C, Yue L, Zhang P. Spin-affected reflexive and stretching separation of off-center droplet collision. Phys Rev Fluids. 2022;7(1):013603.

[29] Laroche C, Bacri J-C, Devaud M, et al. Observation of the resonance frequencies of a stable

torus of fluid. Phys Rev Lett. 2019;123(9):094502.

[30] Erdmanis J, Kitenbergs G, Perzynski R, et al. Magnetic micro-droplet in rotating field: numerical simulation and comparison with experiment. J Fluid Mech. 2017;821(May):266–295.